# Energetics, skeletal dynamics and long-term predictions in Kolmogorov-Lorenz systems

V. Pelino, F. Maimone

*Italian Air Force, CNMCA*
*Aeroporto "De Bernardi", Via di Pratica Di Mare,*
*I-00040 Pratica di Mare (Roma) Italy*

e-mail: pelino@meteoam.it, maimone@meteoam.it

**Abstract**

We study a particular return map for a class of low-dimensional chaotic models called Kolmogorov-Lorenz systems, which received an elegant general Hamiltonian description and includes also the famous Lorenz-63 case, from the viewpoint of energy and Casimir balance. In particular it is considered in detail a subclass of these models, precisely those obtained from the Lorenz-63 by a small perturbation on the standard parameters, which includes for example the forced-Lorenz case in Ref.[6]. The paper is divided into two parts. In the first part the extremes of the mentioned state functions are considered, which define an invariant manifold, used to construct an appropriate Poincarè surface for our return map. From the 'experimental' observation of the simple orbital motion around the two unstable fixed points, together with the circumstance that these orbits are classified by their energy or Casimir maximum, we construct a conceptually simple skeletal dynamics valid within our sub-class, reproducing quite well the Lorenz map for Casimir. This energetic approach sheds some light on the 'physical' mechanism underlying regime transitions. The second part of the paper is devoted to the investigation of a new type of maximum energy-based long-term predictions, by which the knowledge of a particular maximum energy 'shell' amounts to the knowledge of the future (qualitative) behaviour of the system. It is shown that, in this respect, a local analysis of predictability is not appropriate for a complete characterization of this behaviour. A perspective on the possible extensions of this type of predictability analysis to more realistic cases in (geo)-fluid dynamics is discussed at the end of the paper.

## I. INTRODUCTION

A largely studied model exhibiting chaos has been that obtained by Lorenz in 1963 [1] by a drastic truncation of the fluid-dynamics equations as applied to the problem of thermal convection. Since then, Lorenz equations have been derived and applied in a wide variety of contexts [2,3,4,5,6].
Particularly, it was originally thought of for meteorological applications, where low dimensional models of this type ([7], References therein) still form the basis of our current understanding and intuition about atmospheric and climate non-linear dynamics. In this respect, such models have been used [8] to infer the limits of atmospheric predictability, which have been estimated in 10-15 days, at least as far as we are concerned with baroclinic-eddy modes. This conclusion about the existence of a predictability horizon for the atmosphere has been achieved from very different approaches. A remarkable classical argument, for instance, has been based on topological fluid dynamics and was carried out by V.I. Arnold in 1950 [9], who estimated in two weeks this time horizon, before the advent of massive computer capabilities applied in this field.
Nevertheless, it was pointed out that predictability, i.e. error growth, strongly depends on the (space and time) scales considered (see, for example, Ref.[7]). In fact, it turns out that as far as we are concerned with the coarser-grained atmospheric flows generally known as weather regimes, which capture the atmospheric variability comprised between 10 and 100 days, predictability of state transitions is greatly enhanced, as shown for example using statistical autoregressive methods [10].

It is well known that thermo-fluid dynamical equations for ideal systems can be put in an elegant (non-canonical) Hamiltonian form [11,12]. While this formulation obviously adds nothing to the physics of the fluid systems considered, the main mathematical structures turns out to be much more clear in this approach, in particular an infinite class of enstrophy-like Casimir functions are easily identified, whose time variations are only due, eventually, to the action of dissipation and forcing mechanisms.

On the one hand, it is a quite general consequence of the model truncation necessary to solve numerically the equations, like the one employed in the simulation of the atmosphere for operational forecasting purposes (e.g., [13]), that the Hamiltonian form is not preserved in the resulting discrete system of (ordinary) differential equations.

On the other hand, it has recently been shown that an important class of 3-D chaotic models, including Lorenz-63 and a (symmetric) variant of Lorenz-84 models, the so called Kolmogorov-Lorenz systems, indeed can be equipped with a natural geometrical group structure equivalent to that used in the infinite-dimensional case [9], which results in a rigorous Hamiltonian formalism [14, 15].

The paper is organized as follows. In Sec II the Hamiltonian formulation of Kolmogorov-Lorenz system is summarized and the related geometrical (return) maps introduced. Sec. III focuses on the energetics of Lorenz-63, with a discussion of a more general (forced) case (III A), then a skeletal dynamics is constructed, based on this energetics (III B). In Sec. IV the relation between energetics and predictability is investigated, starting with a study of linear instability (IV A), and continuing with a discussion of long-term predictions using energy-maxima (IV B). Sec. V is devoted to a proposal aimed to extend the connection found to more realistic fluid dynamical systems, while Sec. VI concludes the paper with few final remarks.

## II. HAMILTONIAN FORMULATION AND GEOMETRICAL MAPS

The main peculiarity of the model consists in the following: for a suitable choice of the parameters, a chaotic macro-dynamics can be identified, which consists of sudden and unpredictable transitions between two separate regions in phase-space, which will be referred to as the left ($\Psi_L$) and right ($\Psi_R$) regions covering the attractor $\Psi = \Psi_L \cup \Psi_R$.

As it can be easily shown [16], this macro-dynamics can be characterized by a simple statistical law, using a new discrete-time map having a precise physical meaning.

On the other hand, it can be demonstrated that have shown that Lorenz-63 system

$$\begin{cases} \dot{x}_1 = -\sigma x_1 + \sigma x_2 \\ \dot{x}_2 = -x_1 x_3 + \rho x_1 - x_2 \\ \dot{x}_3 = x_1 x_2 - \beta x_3 \end{cases} \quad (1)$$

is not but a particular example of a much wider class of models with a well-defined geometrical interpretation, the so called Kolmogorov-Lorenz equations, in which a clear distinction among Hamiltonian, dissipative and forcing terms is made [14].

To be specific Eq. (1) can be written as Lie-Poisson equations on the algebra of $SO(3)$ spatial rotation group

$$\dot{x}_i = \{x_i, H\} - \Lambda_{ij} x_j + f_i \qquad (i = 1,2,3) \qquad (2)$$

assuming the following gyrostat-like Hamiltonian

$$H = \frac{1}{2}\Omega_k x_k^2 + h_k x_k \qquad (3)$$

with $\Omega = diag(2,1,1)$, dissipation matrix $\Lambda = diag(\sigma,1,\beta)$, an axisymmetric rotor $\mathbf{h} = (0,0,-\sigma)$ and external forcing $\mathbf{f} = (0,0,-\beta(\rho+\sigma))$. An important result of this formalism, is that there is not chaotic behaviour in the system for $\mathbf{h} = \mathbf{0}$.

Here, the brackets represent the algebraic structure of Hamiltonian part of the systems described by function $H$, and the cosymplectic matrix $\mathbf{J}$ [17],

$$\{F,G\} = J_{ik}\partial_i F \partial_k G. \qquad (4)$$

For a conservative system, in the local co-ordinates $x_i$, the Lie-Poisson equations read as

$$\dot{x}_i = \varepsilon_{ik}^j x_j \partial_k H, \qquad (5)$$

where the Ricci tensor $\varepsilon_{ik}^j$ represents the constants of structure of the algebra $\mathbf{g} = \mathbf{so(3)}$ and the cosymplectic matrix assumes the form $J_{ik} = \varepsilon_{ik}^j x_j$; in this formalism $\mathbf{g}$ is endowed with a Poisson bracket characterized by Eq. (2) for functions $F,G \in C^\infty(\mathbf{g}^*)$.

Casimir functions $C$ are given by the kernel of bracket (4), i.e. $\{C,G\} = 0, \forall G \in C^\infty(\mathbf{g}^*)$, therefore they represents constants of motion of the Hamiltonian system, $\dot{C} = \{C,H\} = 0$;

In a geometrical language, they define a foliation of the phase space of Eqs. (3) [9], and its trajectory is given by the intersection of the ellipsoid $\mathrm{E}: \frac{1}{2}\Omega_k x_k^2 + h_k x_k = H$ with the sphere $S^2$ given by the Casimir $x^k x_k = C$, representing the Euler equations for the rigid body,

$$\mathbf{x}(t) = \mathrm{E} \cap S^2. \qquad (6)$$

In the general case of a dissipative-forced system dynamics is also constrained by Eq. (6), but in this case the two geometrical objects $S^2$ and $\mathrm{E}$ do expand and contract in a chaotic way reaching a set of maximal and minimal values during their evolution. In particular, it is natural to consider the set of $\mathbf{x}(t)$ such that $S^2$ reaches a relative maximum or minimum radial value. This is given by imposing the simple condition $\dot{C}(t) = 0$, that using Eq. (2) reads as $-\Lambda_{ik} x_i x_k + f_i x_i = 0$ and defines an invariant ellipsoid, whereas the curve $\ddot{C}(t) = 0$ on this surface separates the regions of maxima and minima. Besides the maximum distance of the points on the ellipsoid from the origin of the coordinates defines, by construction, the radius $R$ of a sphere in which the attractor is entirely contained. Incidentally, these geometrical conditions can be used to verify the consistency of this type of numerical simulations.

Let's now specialize to the Lorenz case of Eq. (1). As it is well known, for a suitable choice of parameters the solutions of Lorenz equations are given by chaotic trajectories revolving around the three unstable fixed points of Eq. (2). In Ref. [16] a clear geometrical method has been established to compute the frequencies of these chaotic oscillations.

After the translation $(x_1, x_2, x_3) \to (x'_1, x'_2, x'_3) = (x_1, x_2, \sqrt{\rho}x_3 + \beta(\rho+\sigma)/2\sqrt{\rho})$ we can write the condition $\dot{C}(t) = 0$ in a geometrical formalism as the following ellipsoid of rotation

$$E_C : \sigma x'^2_1 + x'^2_2 + x'^2_3 = \left[\frac{\beta(\rho+\sigma)}{2\sqrt{\rho}}\right]^2, \qquad (7)$$

that is a fixed manifold with respect to the chaotic motion on the attractor $\mathbf{x}'(t) \in \Psi$, with a fractal structure for $\Psi$. In this way it is simple to choose the Poincarè map for Lorenz attractor as the intersection

$$x'(t_k) = E_C \cap \Psi \quad . \qquad (8)$$

Such intersection defines four symmetric regions on the ellipsoid $E_C$ representing the sets of $max(C)$ and $min(C)$ for the two lobes of $\Psi$, as shown in Fig. 1. Return maps for these points give the periods of rotation and revolution, i.e. respectively about each lateral fixed point and about the central unstable point.

To fix the ideas let us concentrate on the first return map of one of the maxima, the left one, say. The spectrum of the return times depicted in Fig. 3 shows a definite band structure with a constant spacing separation, which can be interpreted as follows: the first band (index $n=0$) represents a simple rotation in the left lobe, the second ($n=1$) corresponds to a semi-rotation in the left lobe followed by a complete rotation in the other lobe plus a semi-rotation in the starting lobe, etc.

So the $n$-th band corresponds to n-complete rotations in the right lobe and return.

Besides a characteristic time $\tau_0 \approx 0.66$ units, defined by the constant spacing, emerges.

Because of the reflection symmetry $x_i \to -x_i$, $i=1,2$ the same behaviour can be verified (at least in a statistical sense) taking as starting point the maximum on the right side, while a slightly different statistics is obtained for the minima.

It should be stressed that even if a similar result can be expected also for other suitably chosen Poincarè sections, the regions of extrema on the invariant ellipsoid is a very natural choice, being automatically defined for all the models within our class. Also, we point out the identical structure of Lorenz map for the Casimir function relatively to the standard $x_3$ Lorenz map; moreover in order to distinguish the jumps of $max(C)$ between $\Psi_L$ and $\Psi_R$, it could be more natural to consider the new map for the set of values $D_n(\mathbf{x}) = \sigma(x_1) \cdot C_n(\mathbf{x})$, where $\sigma(x_1) = \pm 1$ represents the region of the attractor where the discrete dynamics happens. In the following we will consider a still different map, which in any case will be explicitly regime-sensitive.

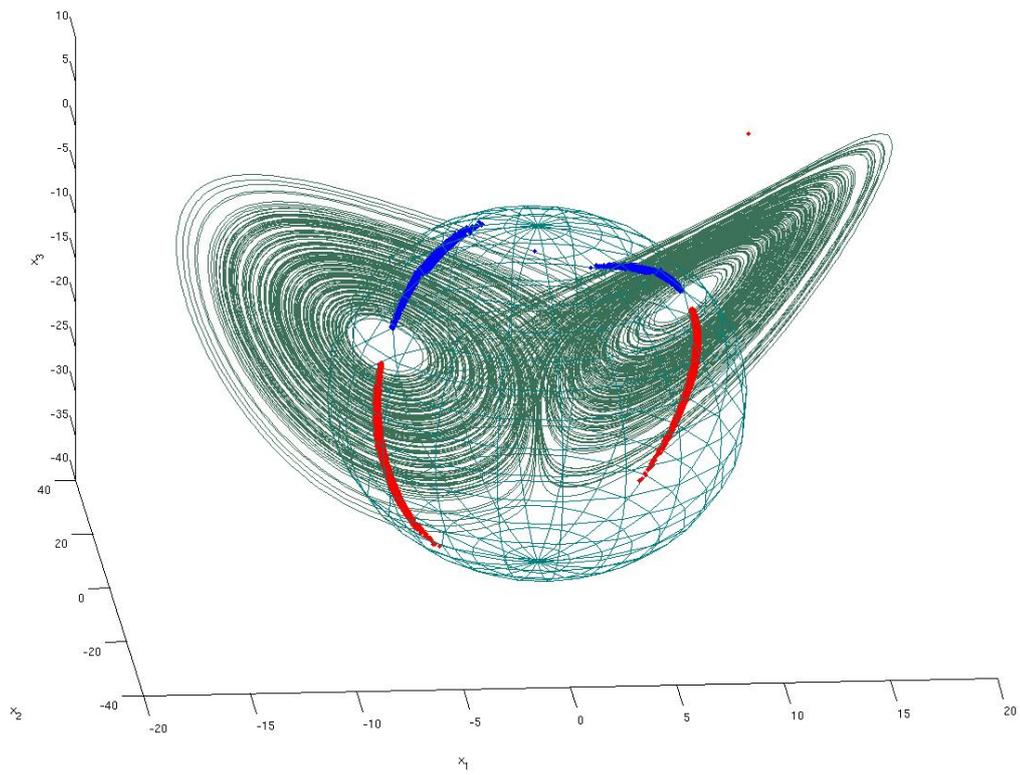

FIG. 1. Ellipsoid of Casimir extremes intersecting the attractor in the four regions, the two ones behind representing the set of maxima, and the remaining two the minima.

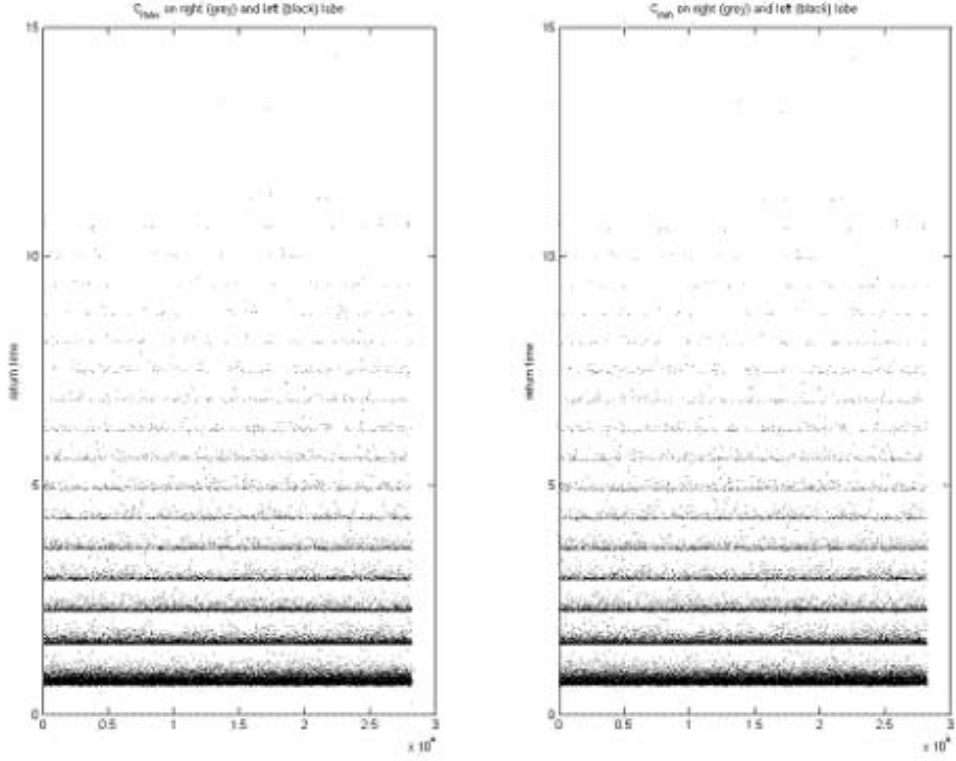

FIG. 2. Band structure of the return times.

## III. ENERGETICS OF LORENZ-63

We'll consider now the behaviour of the system from the viewpoint of the Energy and Casimir functions.

### A. Energy balance between forcing and dissipation

It is a general statement that even if energy is not conserved, the energy gain in a given finite trajectory on the attractor should be equal to the time integral of the sum of the injection power and of the dissipation power along the track,

$$E(t_2) - E(t_1) = \int_{t_1}^{t_2} \dot{H} dt = -\int_{t_1}^{t_2} \Lambda_{kj}(\Omega_k x_k + \omega_k) x_j dt + \int_{t_1}^{t_2} f_k(\Omega_k x_k + \omega_k) dt. \tag{9}$$

Since (natural) forcing injection and dissipation vary in a different way on the attractor, i.e. they are functions, respectively, linear and quadratic of the phase-space coordinates, this fact constrains the representative point on the attractor to follow well-defined paths.
The three curves depicted (not in scale) in Fig. 3 represent the total energy (bottom curve), the forcing-dissipation power (highest amplitude oscillations in the figure) and the $x_1$ coordinate. From the analysis of these curves it emerges that from an energetic viewpoint the system, soon after a passage from the companion lobe, trajectory tends to acquire energy around the unstable central fixed point, where the (positive) forcing pushes the systems towards higher energies, whose maximum is reached at a peripherical side of the lobe, as it is possible to see from the Fig. 4, in which energy injection per unit length is represented, i.e. the quantity

$$\dot{E}/\|\mathbf{v}\|, \qquad (10)$$

where $\mathbf{v} = (\dot{x}_1, \dot{x}_2, \dot{x}_3)$ is the velocity field.

After the energy maximum has been reached, dissipation becomes more and more important along the system's trajectory, causing energy loss. If the energy loss exceeds a given threshold then the energy gain near the centre of the attractor is not sufficient to sustain another oscillation around the fixed point of the current lobe, and the system is '*pushed*' into the other regime. Otherwise the system is allowed to experience another turn within the same lobe along a more external trajectory, and so on. It is then possible that the system's state performs $n-$turns within a given lobe, experiencing at each successive step more and more external trajectories around the fixed point. As it is shown in Fig. 5, because of greater energy gain near the attractor's centre, it happens that very internal trajectories acquire much less energies when passing there than more external ones, and at the same time they suffer only for a small dissipation in a turn. As a result the small amount of energy acquired allows for a small radius increment of the trajectory itself in a complete turn. This fact, on the other hand, allows for a high number of consecutive turns inside the given lobe, up to the 'critical' one, in which dissipation becomes unsustainable and the system 'decades' in the other regime. This is the qualitative explanation of the cusp form of the Lorenz map from an energetic viewpoint. Incidentally it can be verified that the same type of maps can be obtained for both energy and Casimir.

Moreover, it results from a simple inspection of Fig 3, that the most external orbits above to the critical radius end up in very internal orbits on the other lobe, which means a large number of consecutive turns within this latter. This circumstance strongly suggests that maximum energies are linked to the number of turns the system will perform on the other lobe, i.e. to the long-term behaviour of the system.

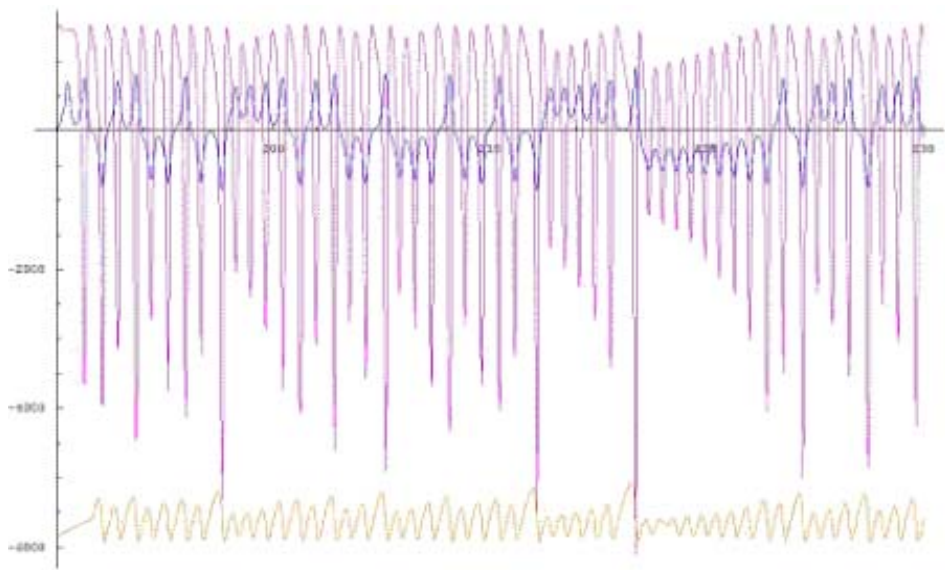

FIG. 3. Time variation of total energy (bottom curve), forcing-dissipation power (highest amplitude oscillations in the figure), and $x_1$ coordinate. The variables are represented not in scale, in particular the plotted functions

are respectively: $E(t) - 6000$, $\dot{E}(t)$, $50 * x_1$.

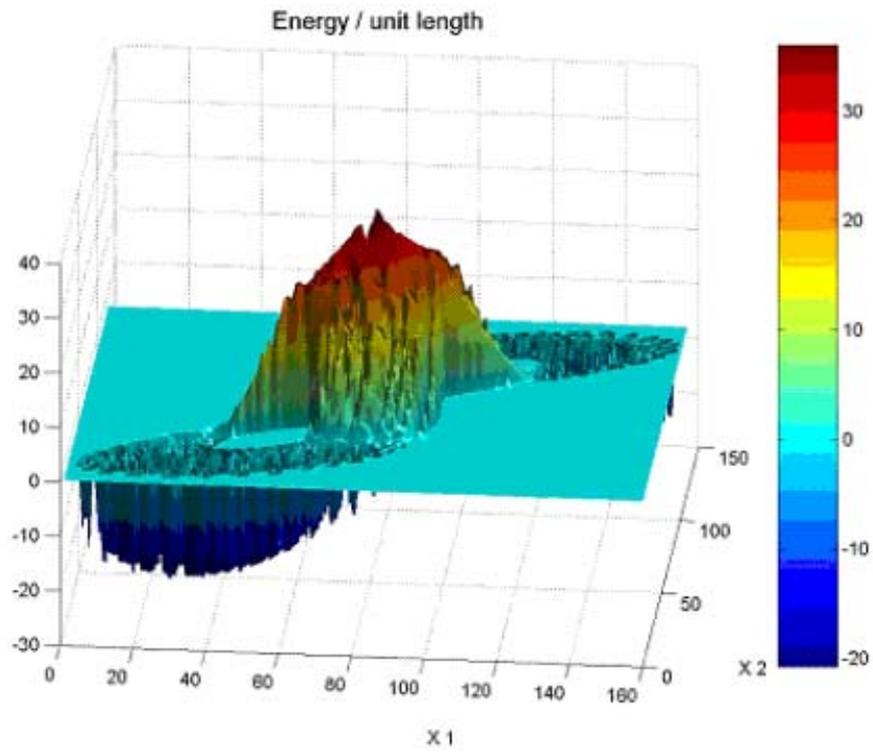

FIG. 4. Energy acquired per unit length represented in the plane $(x_1, x_2)$ by time averaging on the $x_3$ direction.

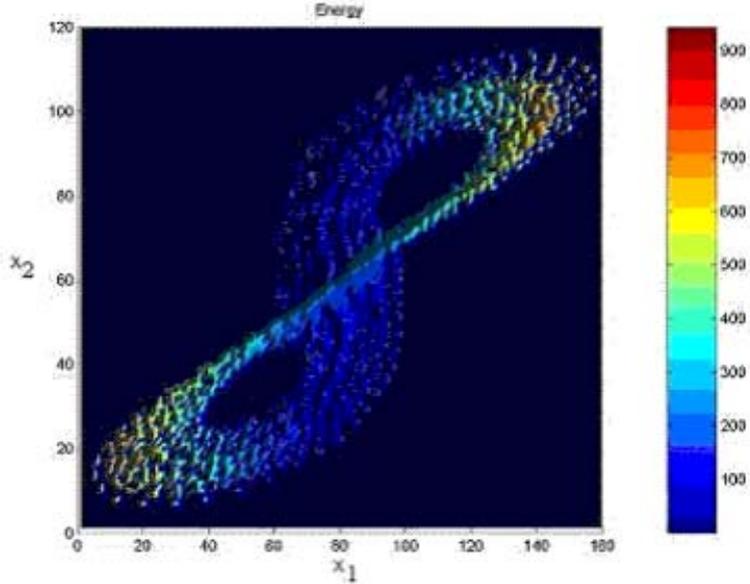

FIG. 5. Total energy represented in the plane $(x_1, x_2)$ by time averaging on the $x_3$ direction.

Finally, in order to give the flavour of the generality of the present approach, we consider briefly the case studied in Ref.[6], in which a regime-selective behaviour is obtained through the introduction of a further (weak) forcing in the $x_1$ and $x_2$ equations, parametrized by the angle $\vartheta$. In a more general case, assuming a forcing **f**

$$\begin{cases} f_1 = f \cos\varphi \cos\vartheta \\ f_2 = f \cos\varphi \sin\vartheta \\ f_3 = f \sin\varphi \end{cases} \quad (11)$$

where $f = -\beta(\rho + \sigma)$, it is easy to show that for $\varphi \neq \pi/2$ there is a symmetry breaking in the Lorenz equations leading to different statistics and predictability in the two lobes. What happens, from an energetic viewpoint, is that the introduction of this additional forcing modifies the central 'hill' of Fig. 7 (with a simultaneous small displacement of the lateral unstable points, the form of the attractor remaining essentially the same), in such a way that it becomes asymmetric, i.e. a bit more pronounced, with the values set in Ref.[6], on a lobe than on the other, the dissipation term remaining the same. In such a way, on the lobe in which it is more pronounced, internal orbits are forced to acquire more rapidly energy, so that the representative point is 'pushed', on the average, a bit faster towards the other regime. Conversely, for an internal orbit originating in the opposite lobe, maximum-energy growth is decelerated, so that the (mean) residence time of the representative point in that lobe is increased. This gives rise to the observed difference in the PDF on the two lobes.

**A. Orbits and skeletal dynamics**

It is an 'experimental' observation that, starting from the Lorenz type of the Kolmogorov-Lorenz systems with the original Lorenz parameters, and for a wide range of variations around them (in the chaotic regime), including the case of Ref.[6], that some peculiar properties of the dynamics are robust. On the one hand, in fact, the fractal dimension preserves approximately the value $d \approx 2.06$. On the other hand, macro-dynamics remains qualitatively unchanged.

Moreover, it is important to note that the sets of maximum energies, at variance for example with those of minimum ones, have a natural order, i.e. maximum energies are growing functions, in each lobe, of the orbit radius (see Fig. 6).

In this case, apart from the fine-grained structure, the fractal attractor is geometrically approximated by a two-dimensional manifold. To be specific we consider two distinct surfaces, one for each wing of the attractor, which glue together at the points in which the lines, originating in a given region, enter the opposite wing. We obtain such surfaces by interpolating $10^3$ points for each lobe, and uniformly distributed over them, with two third-order polynomials in the variables $x_1$ and $x_2$, $z(x_1, x_2)$. So the left and right wings are geometrically described respectively by the metrics (induced by the usual Euclidean 3-D metric of the phase-space) $g_L$ and $g_R$.

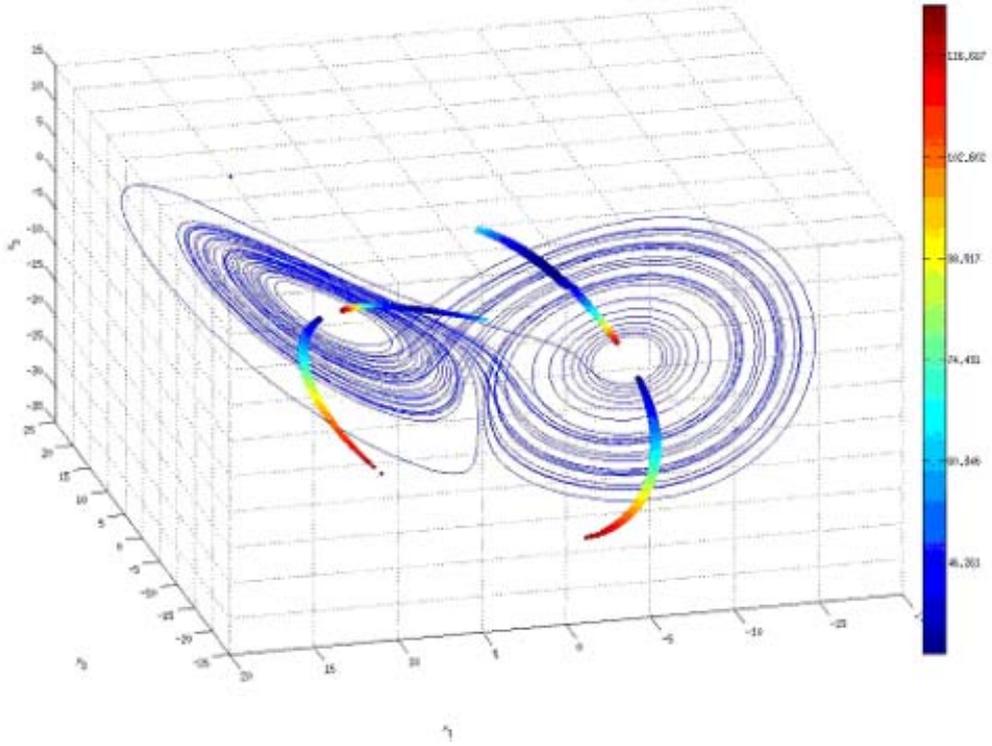

FIG. 6. The ordered sets of Casimir maxima (behind) and minima are displayed on the attractor. At variance with the minima, maxima are ordered, growing from the internal orbits to the external ones.

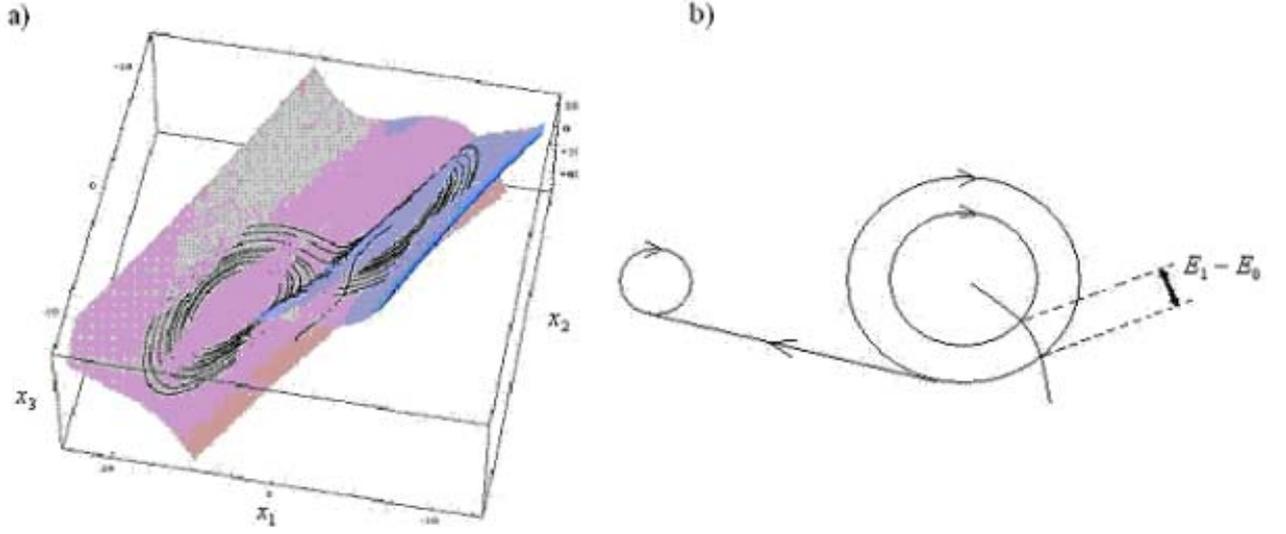

FIG. 7. (a) Connecting surfaces obtained by interpolating $10^3$ points for each wing using two third-degree polynomials in the variables $x_1$ and $x_2$: $A + Bx_1 + Cx_1^2 + Dx_1^3 + Ex_2 + Fx_1x_2 + Gx_1^2x_2 + Hx_2^2 + Ix_1x_1^2 + Lx_2^3$, with
$A = -22.95; B = 5.25; C = -0.62; D = 0.03; E = -3.31; F = 0.76; G = -0.05; H = -0.21; I = 0.03; L = 0.01;$
for the right side, and the polynomial obtained from this latter by reversing the coordinates for the left side (b) Schematic view of the orbits intersecting the curve of energy maxima.

In this setting, of course, the fractal regions of points of maximum energy are reduced to lines, being the intersections between the attractor surfaces and the invariant energy ellipsoid.
On the basis of the above 'experimental observations' we construct our skeletal dynamics by making the following assumptions:

   i)   Dynamics takes place on a two-dimensional manifold, union of the two surfaces described above;

   ii)  the basic geometrical elements of the trajectories are: 1) circles (in the metric $g$) around the two lateral fixed points; 2) lines connecting circles on the two lobes;

   iii) each circle, or orbit, intersects the curve of the energy maxima at a certain point. Then the next orbit is defined via the energy (forcing +dissipation) acquired in a complete turn, in such a way that it will intersect the curve at a new point whose energy corresponds to the acquired energy added to the previous maximum (see Fig. 7(b));

   iv)  when the energy acquired within one orbit happens to be negative, the orbit itself is not completed, and at a certain point the motion is switched to a tangent line, until it reaches the other wing; from this moment on, the motion continues on the orbit passing through the crossing point, and the cycles restart on the other 'wing'; the switching point is determined in a self-consistent way by requiring that the energy acquired along the segment connecting the two lobes amounts to the energy difference between the end points.

From these rules we will derive a number of consequences.

We start with the 'forecast' of the number of turns the representative point will experience within the starting lobe and in the successive one soon after we have detected a energy maximum $E_0$ at a certain instant.

The number of turns made within the starting lobe is the maximum $n$ such that $\dot{H}/\|\mathbf{v}\| \geq 0$, while the maximum energy acquired in the lobe L( R) is given (in an explicit) form by

$$E_n = E_0 + \sum_{i=1}^{n} \oint_{C_i(C_{i-1})} \frac{-\Lambda_{kj}(\Omega_k x_k + \omega_k)x_j + f_k(\Omega_k x_k + \omega_k)}{\sqrt{\sum_{i=1}^{3}[\{x_i, H\} - \Lambda_{ij}x_j + f_i]^2}} ds, \quad (12)$$

where

$$ds = \sqrt{g_{ij}^{L(R)}(x_1, x_2) dx_i dx_j}, \qquad g_{i,j}^{L(R)} = \begin{pmatrix} [1+\partial_{x_1}z]^2 & (\partial_{x_1}z)(\partial_{x_2}z) \\ (\partial_{x_1}z)(\partial_{x_2}z) & [1+\partial_{x_2}z]^2 \end{pmatrix}. \quad (13)$$

Using the above 'theory of orbits', we compute the successive (maximum) energy levels starting from a given maximum. The result is depicted in the Fig. 8, against the true cusp-map of maxima.

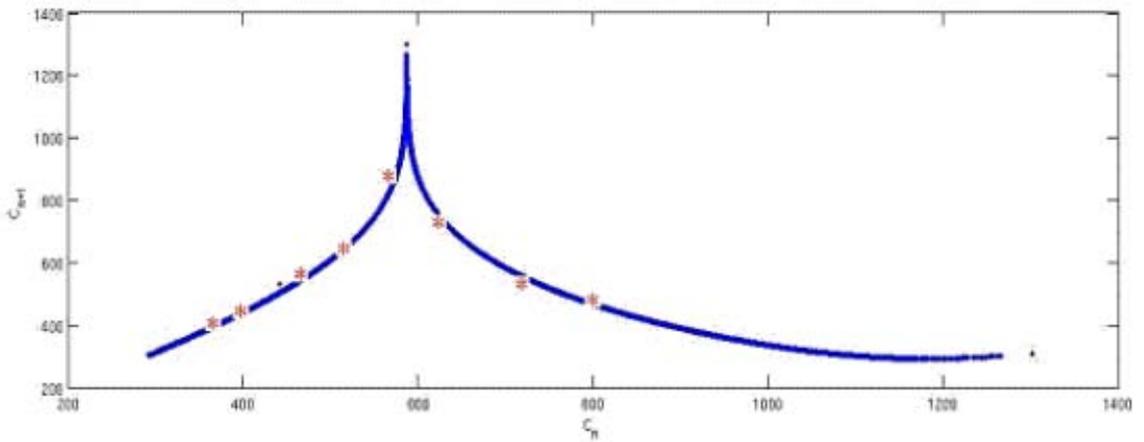

FIG. 8. Lorenz map points (dots) for Casimir plotted against the experimental map (continuous line).

Given the good correspondence of the skeletal model points with the Lorenz map, we can conclude that our theory of orbits captures the gross features the dynamics. Since each orbit is, by definition, unambiguously associated with a maximum of energy, we can conclude that this quantity is a good control parameter for the system behaviour.

It should be stressed that the present skeletal model is only intended to give an energetic justification of the Lorenz map, i.e. for a given Casimir maximum it furnishes the approximate next value of this latter, the full chaotic dynamics remaining, in any case, not reproducible by substituting the fractal set of the attractor with a regular surface.

## IV. ENERGETICS AND PREDICTABILITY

In this section we discuss the connection between energetics and predictability. We analyze this latter firstly in the 'local' (linear) approach, then we pass to a longer term predictability analysis based on nonlinear evolution, and to an interesting use of energy and Casimir maxima as control variables to infer qualitative future behaviour of the system.

### A. Local instability by linear perturbation analysis

Let us calculate the analytic expression of the perturbation growth rate in the natural Euclidean metric as a function of coordinates, depending parametrically on the vector of the initial perturbation and on the (small) time interval of the evolution considered.

To be specific, we consider the evolution of two nearby initial points $x_i^0$ and $\tilde{x}_i^0$. The dynamical equation for the difference is

$$\frac{d}{dt}(x_i - \tilde{x}_i) \equiv \frac{d\delta x_i}{dt} = \{\delta x_i, H\} - \Lambda_{ij}(\delta x_i), \qquad (14)$$

from which, considering a sufficiently small time interval $\Delta t$, we get equations for $\delta x_i(\Delta t)$

$$\begin{aligned}\delta x_i(\Delta t) &\approx \delta x_i^0 + \Delta t \left[\Omega_k \varepsilon^j{}_{ik}(x_k x_j - \tilde{x}_k \tilde{x}_j) + \omega_k \varepsilon^j{}_{ik} \delta x_j(\Delta t) - \Lambda_{ij} \delta x_j(\Delta t)\right] \\ &\equiv \delta x_i^0 + \hat{M}_{ij} \delta x_j(\Delta t).\end{aligned} \qquad (15)$$

Solving this linear system with respect to $\delta x_i(\Delta t)$, the perturbation vector growth rate can be expressed conveniently by

$$g = \frac{1}{\Delta t} \ln \frac{\sum_{k=1}^{3} \delta x_k^2(\Delta t)}{\sum_{k=1}^{3} \delta x_k^{0\,2}} = \frac{1}{\Delta t} \ln \frac{\sum_{k=1}^{3} \left(\det\|\hat{M}_{ij}^{(k)}\| / \det\|\hat{M}_{ij}\|\right)^2}{\sum_{k=1}^{3} \delta x_k^{0\,2}}, \qquad (16)$$

where matrix $\|\hat{M}_{ij}^{(k)}\|$ is obtained from $\|\hat{M}_{ij}^{(k)}\|$ by substituting the $k$-th column with the vector $(\delta x_1^0, \delta x_2^0, \delta x_3^0)^T$.

In the graphics below it is depicted the growth rate along the curve of maximum energies of the skeletal attractor setting a time-step $\Delta t = 0.02$ and choosing the initial errors $\delta x_i^0 \approx O(10^{-1})$ respectively along the curve and in the fixed direction $(1,1,1)$. The result is that there is only one evident maximum, corresponding to an energy E and a Casimir C, for which local predictability reaches its minimum. This point identifies precisely the critical point of regime transition. Rapid oscillations in the second case are due to the fact that one of the initial conditions in general does not belong to the attractor. Nevertheless the long-term predictability we will be concerned with, cannot be identified by a linear analysis of instability such as bred-vectors or singular values methods, because, as we will see in the following, the knowledge of the starting energy shell above the critical point identifies precisely the number of turns, and then the mean time the system will remain, in the opposite regime.

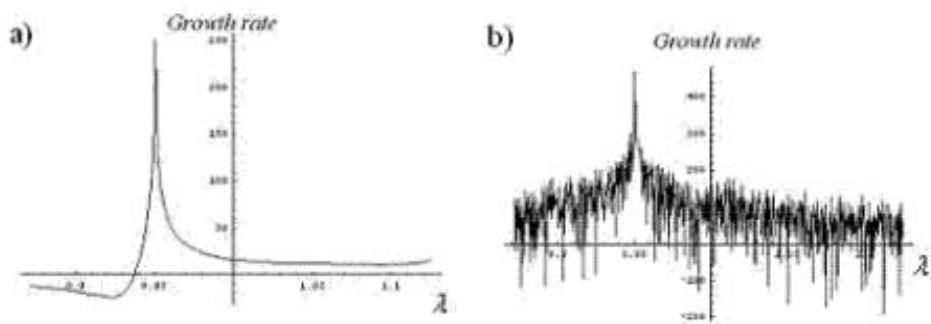

FIG. 9. Growth rate calculated along the curve of maxima (parametrized by the affine parameter λ) going from from the internal orbits (lower values of λ) to the external ones, and taking the initial perturbations along the curve(a) and in the (1,1,1) direction in the phase-space(b).

## B. Long-term predictions using energy maxima

We have already seen and interpreted the band structure of the return times of our Poincarè map. It is interesting that, considering the return times as a function of the energy or of the Casimir, we find the ordered structure depicted in Fig. 10. Starting from an energy maximum on a given lobe, the system's state will wander around through the attractor and then, after a certain time, it will return to another maximum on the same lobe. The map 5.1 contains the return times plotted against the Casimir as obtained in a long-time integration of $40 \times 10^6$ calculation steps. The emerging band structure of the Casimir (the same holds true for the energy) can be easily interpreted in the following way: the first band corresponds to a turn within the starting lobe, while the $(n+1)$-th band corresponds to $n$-turns on the other lobe and return.

So, in other words, at a certain Casimir range it corresponds a precise return time, so that the knowledge of this quantity, according to the discussion above, amounts to the knowledge of the number of turns on the other lobe.

An interesting question is whether a given Casimir maximum also determines its next map iteration value, in such a way that one could be able, not only to forecast the next number of turns, but also the successive ones, and so on, encoding a complete symbolic dynamic description of the system in an ordered sequence of number of turns, alternatively performed on the two lobes.

In other words, we can look at our system as a black box, whose input is the entrance energy, i.e. an initial maximum of energy on a given lobe, and the output is the energy maximum on the same lobe when the system returns back. In Fig. 11 it is depicted the map of Casimir at a given lobe, for example the right one, $C_{n+k}^r = f(C_n^r)$. It's immediate to note that to a little uncertainty $\Delta C_n^r$ in the entrance Casimir it corresponds, in general, a greater uncertainty $\Delta C_{n+k}^r$ in the exit Casimir. Moreover the ratio $\Delta C_{n+k}^r / \Delta C_n^r \approx \partial f / \partial C_n^r$ is, on the average over the bands, a growing function of the band index up to $i = 10$.

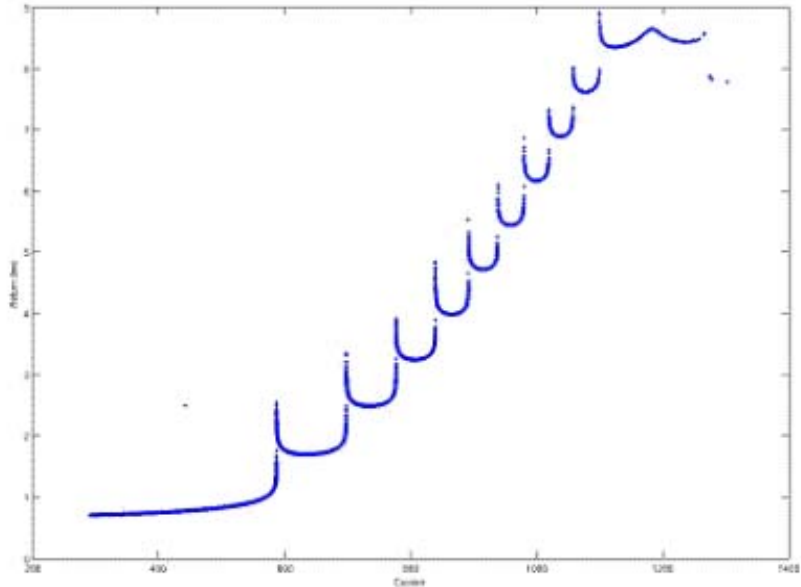

FIG. 10. Return times as a function of Casimir

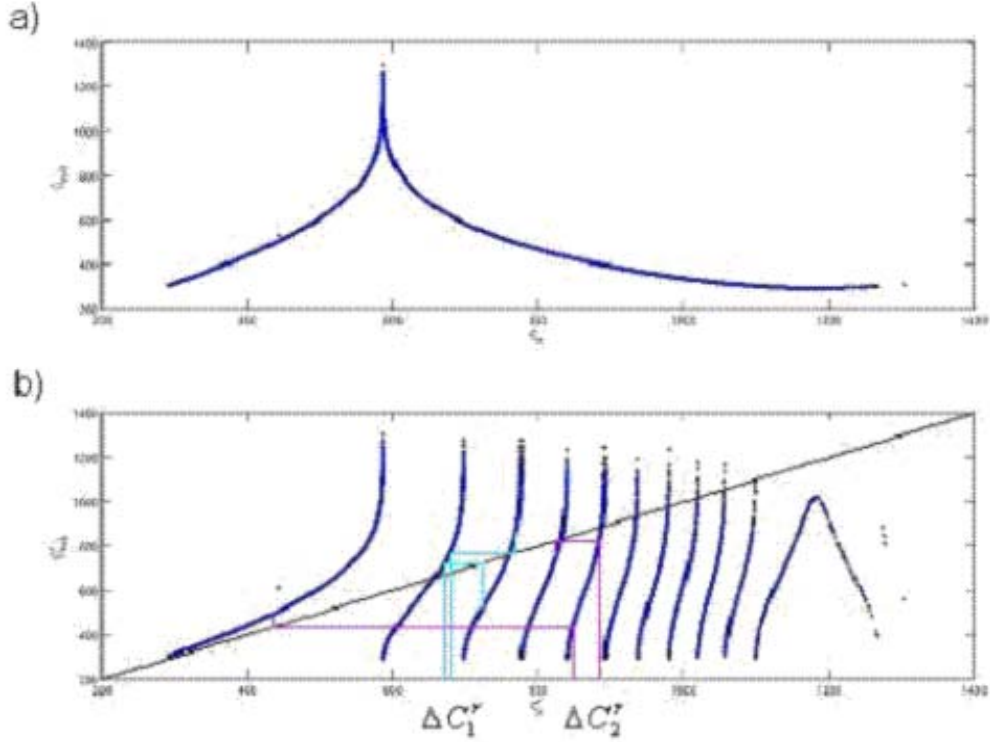

FIG. 11. Lorenz map for Casimir (a); Casimir maximum on the right side $C^r_{n+k}$ as a function of the previous Casimir maximum on the same side after $k-$turns $C^r_n$ (b).

In Fig. 11(b) two examples are considered. In the first, starting with the uncertainty $\Delta C^r_1$ of the right lobe maximum, the system comes back at the next passage with a greater uncertainty, but with the Casimir range extremes still lying on the same band (the third one), meaning that after the subsequent jump the number of turns is still predictable.

In the second example, instead, the uncertainty $\Delta C^r_2$ propagates in such a way that after the first passage the number of turns on the other lobe is precisely determined to be 4, but after the following transition this number is uncertain, ranging from 0 to 3.

Now, taking the set of maximum-energy points comprised within the extremes of a band, we can consider the flux-tube determined by dynamics in a neighbourhood of this set. We will refer later on to these (local) fux-tubes as to energy or Casimir shells.

We will now perform an ensemble long-term prediction to test, in particular, the sensitivity of a region enclosed in a given energy shell to the future behaviour of the system, in comparison with the sensitivity of somewhat other region on the attractor having the same phase-space extension (in the natural Euclidean metric). Alternatively we could have used the metric defined by energy and Casimir themselves.

To be specific, we will consider a series of numerical experiments. In each one we take two sets of points, both belonging to the attractor and contained in the intersection of two equal spherical shells; the former contained within a given Casimir shell, the latter centred about a random point on the attractor. Then we let the two sets evolve. The intersection of the two spherical shells with the quasi-2D attractor approximately define two coronae, and we choose the internal and external radius in such a way that the first corona is entirely contained within a shell. The external radius, on the other hand, is somewhat arbitrary, and it is only justified by the need of avoiding, as far as possible, the 'noise' due to very nearby points, which are expected to have in a natural way a similar future behaviour in our comparison experiment.

One such experiment is illustrated in Fig. 12 and 13, respectively for the case of an initial set chosen in a shell of maximum Casimir and in a random point on the attractor. It can be noticed that in the first case all the trajectories, after the transition, perform 5 turns on the other side, while in the second case they diverge rapidly. In the latter case this fact can be accounted for by noticing that the first maximum Casimir associated with the different trajectories have a wide spreading of values.

Here the operating principle is that two-nearby trajectories in the bulk of a lobe, the most densely occupied and internal part of it, can belong to really different histories, arriving within the lobe at maximum energy values corresponding to far distant energy shells. Otherwise, in the opposite case, they may belong to trajectories ending in the same shell, this latter condition ensuring a similar future behaviour.

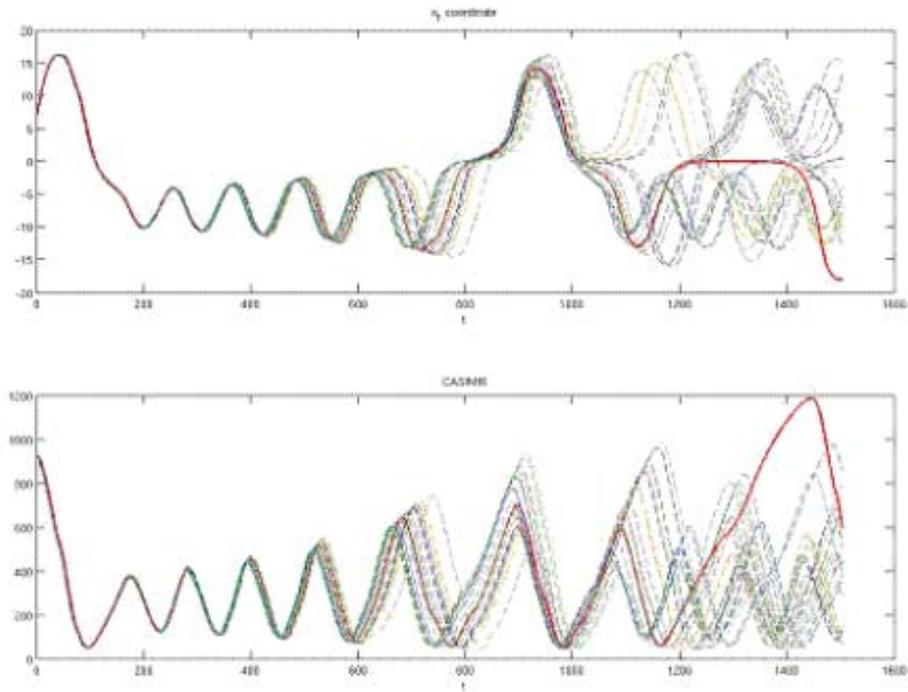

FIG. 12. Evolution of $x_3$ (upper panel) and Casimir (lower panel) starting from several initial conditions, different from each one but belonging to the same Casimir shell.

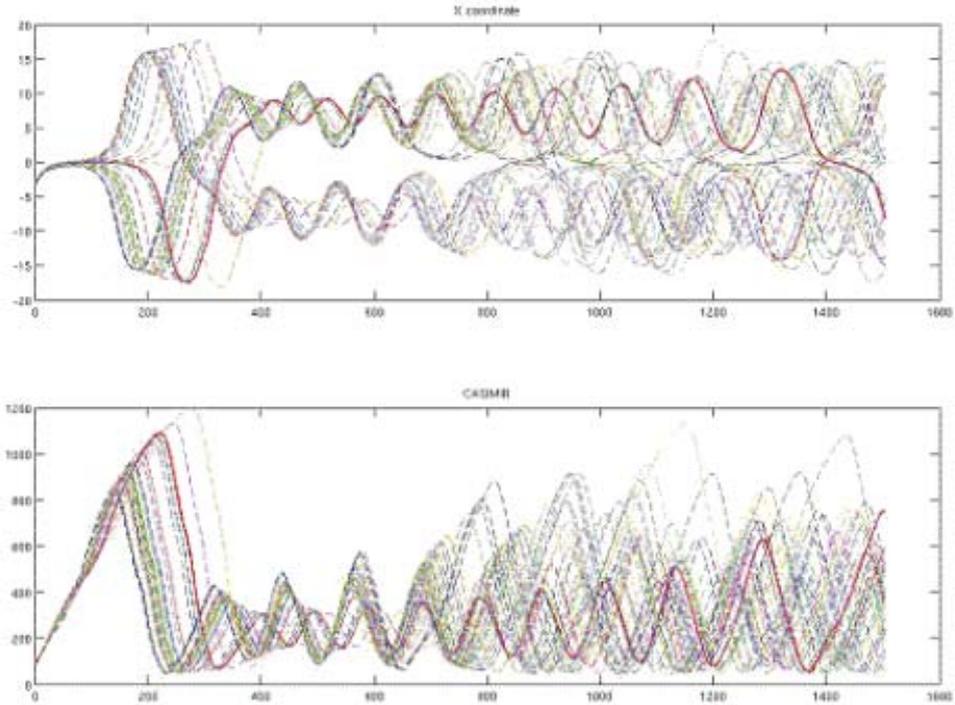

FIG. 13. Evolution of $x_3$ (upper panel) and Casimir (lower panel) starting from several initial conditions chosen within a corona of the same extension as before (Fig. 11), but centred at a random point on the attractor.

It should be stressed that there are cases in which the random corona considered on the attractor results in a good predictability, i.e. all the trajectories are found to experience the same number of turns after the regime transition, and even later, after the subsequent transition, the predictability of this number can be also quite good. In Fig. 14 it is shown such a case, choosing a different point on the attractor. Again we can see that predictability is controlled by the Casimir maximum. In fact the first maximum shows a very small dispersion around a relatively low value, and as a consequence most of the trajectories experience 4 tours in the negative-x regime, while the remaining ones only 3. Things began to be worst with respect to predictability in the subsequent jump, which can be foreseen by the great dispersion of the previous Casimir maximum in the high values range. Still we conclude that future behaviour is better foreseen at the particular moment in which the system reaches its maximum values of the control Casimir functions.

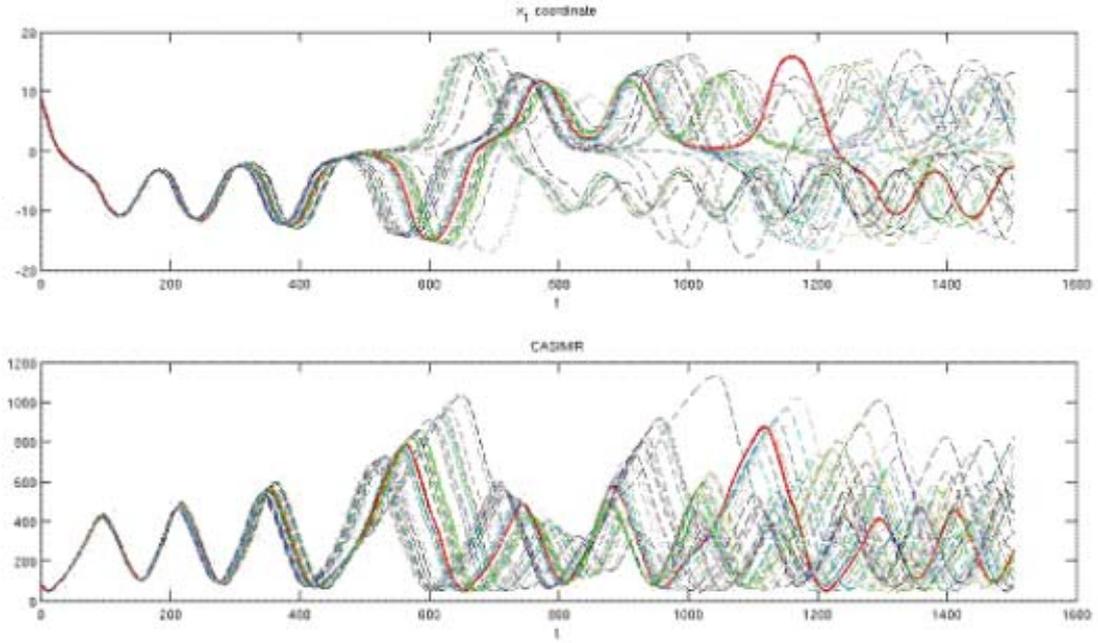

FIG. 14. Evolution of $x_3$ (upper panel) and Casimir (lower panel) starting from several initial conditions chosen within a corona of the same extension as before (Fig. 12) and centred at a different random point on the attractor.

## V. PERSPECTIVES ON THE POSSIBLE EXTENSIONS TO MORE REALISTIC FLUID - DYNAMICAL SYSTEMS

The Hamiltonian description of the Kolmogorov-Lorenz systems and the construction of the invariant ellipsoid of maximum energy and Casimir, with the associated return map, should be easily generalized, at least formally, to the case of a real fluid. Consider, for example, a 3-D homentropic fluid in a given spatial domain. In the ideal case the fluid dynamics would be completely described by the Hamiltonian

$$H[u,x] = \iiint da \left\{ \frac{1}{2}u^2 + \varepsilon\left(\frac{\partial x}{\partial a}, \eta(a)\right) + \Phi(x) \right\}, \qquad (17)$$

where $a$ is an arbitrary labelling of the fluid particles satisfying the only condition that $da = d(mass)$, $\varepsilon$ the internal energy and $\Phi$ an external potential acting on the fluid, $\eta$ is the specific entropy [Salmon book]. The presence of a forcing and a dissipation can be accounted for by adding in the equation of motion the correspond terms

$$\begin{cases} \dfrac{Du}{Dt} = -\dfrac{\delta H[u,x]}{\delta x} \\ \dfrac{Dx}{Dt} = \dfrac{\delta H[u,x]}{\delta u} \end{cases} + Forcing + Dissipation, \qquad (18)$$

Now, the equation $DH/Dt = 0$ defines an hypersurface in the (infinite-dimensional) space of states, whose intersection with the global attractor contains the set S of points of maximum energy. This set should be divided into subsets, one for each possible regime, which in the toy model of Lorenz are easily identified with the fractal set on the two lobes. In the general case regimes are not clearly defined[18,10]. Since our approach is inevitably empirical and aimed to a possible practical application of energy-based long-term predictability to more realistic situations, the problem of defining quasi-stationary regimes can be afforded in an empiric way by considering a discrete grid in the space domain, and performing the usual EOF analysis, as it is currently applied in meteorology[19]. This analysis extracts a certain, hopefully small, number of preferred configurations capturing the gross of the time-variability of the fields. For simplicity the analysis can be restricted to a fixed height or pressure level. Therefore, imagine that the fluid is the atmosphere, and the domain of integration the whole spherical surface of the Earth. A good pressure level to study preferred configurations is the 500-hPa level, and, as far as we are concerned with climatic regimes, we can consider the monthly-averaged geopotential height anomalies. Denoting $\phi'_{i,k}$ the height anomaly for the i-th grid point and the k-th month, the preferred patterns, also called loading-vectors, are obtained by taking the (first) orthonormal eigenvectors $\vec{E}_i$ of the covariance matrix

$$\hat{R}_{i,j} \equiv \frac{1}{N} \sum_{k=1,...,N} \phi'_{i,k} \phi'_{j,k}, \qquad (19)$$

and multiplying it by the corresponding eigenvalue $\lambda_i$ representing the fraction of the total variability [19]. A given height anomaly field can then be written as a linear combination

$$\vec{\phi}'_k = c_{1,k} \lambda_1 E_1 + c_{2,k} \lambda_2 E_2 + .... + \zeta, \qquad (20)$$

where $\zeta$ indicates the residual state not represented by the first preferred patterns. Now, in order to confine the maximum energy values, which can be extracted directly from the observed time series and with the time resolution of the meteorological analysis, to a given regime, it is sufficient to retain only those values whose corresponding coefficient $c_{h,k}$ verifies the condition $f < c_{h,k} < 1$, with $f$ a reference fraction.

Having defined the set of maximum energy points for a given regime, it can be experimentally verified if a given range of maximum energies is associated with a qualitative future behaviour, such as the transition to some other regimes and a subsequent permanence within them for a certain mean time.

It should be stressed that, of course, there is not any *a priori* reason why this correlation should be found in more realistic cases as in our toy models, but, if so, an experimental mapping of these maximum energy ranges and the associated long-term behaviour could become a powerful predictive instrument to forecast regime transitions, for example in realistic fluid systems. As to the possible application to climate regimes, it has been tacitly assumed up to now the presence of a constant forcing. In the case of climate we have of course a periodic forcing due to solar radiation, and the maximum of energy reached in a given regime strongly depends on the period of the year considered, so that this situation requires a special treatment, even within the simple low-dimensional systems we have considered.

## VI. CONCLUSIONS

In conclusion, the use of the Hamiltonian formalism applied to the Lorenz model leads to an interesting 'physical' interpretation in terms of discretized orbits and (maximum-)energy levels. The resulting skeletal dynamics, which takes place on a simplified two-dimensional manifold, seems to capture the basic mechanism underlying dynamics, and, in particular, regime transitions.

Besides, we have classified up to 10 maximum-energy shells in such a way that, starting from a point within one of them, the number of turns the system experiences in the other regime is determined.